\documentclass[prd,groupedaddress,superscriptaddress,nofootinbib,floatfix,preprintnumbers,tightenlines,11pt,longbibliography]{revtex4}
\usepackage[paperwidth=210mm,paperheight=297mm,centering,hmargin=2.4cm,vmargin=2.4cm]{geometry}

\usepackage{float}
\usepackage{nicefrac}
\usepackage{slashed}
\usepackage{mathtools}
\usepackage{amsfonts} % AMS
\usepackage{amssymb} % AMS
\usepackage{amsmath} % AMS
\usepackage{graphicx} % Include figure files
\usepackage{subfigure} % Include figure filesa
\usepackage{array} % array
\usepackage{dcolumn} % Align table columns on decimal point
\usepackage{bm} % bold math
\usepackage{latexsym} % latex symbols
\usepackage{longtable} % long tables
\usepackage{hyperref} % hypertext links 
\usepackage{verbatim}
\usepackage{epsfig}
\usepackage{color}
\usepackage{xcolor}
\usepackage{cancel}
\usepackage{braket}
\newcommand*\dif{\mathop{}\!\mathrm{d}}
\newcommand{\M}[0]{{\mathcal{M}}}
\newcommand{\T}[0]{{\mathcal{T}}}

\newcommand{\nn}[0]{\nonumber}

\begin{document}

\title{
A model-independent framework for determining\\
 finite-volume effects of spatially nonlocal operators
}
%%%%%%%%%%%%%%%%%%%%%

%%%%%%%%%%%%%%%%%%%%%
\author{Ra\'ul A.~Brice\~no}
\email[]{rbriceno@jlab.org}
\affiliation{Thomas Jefferson National Accelerator Facility, Newport News, Virginia 23606, USA}
\affiliation{ Department of Physics, Old Dominion University, Norfolk, Virginia 23529, USA}
%%%%%%%%%%%%%%%%%%%%%
\author{Christopher J. Monahan}
\email[]{cjmonahan@wm.edu}
\affiliation{Thomas Jefferson National Accelerator Facility, Newport News, Virginia 23606, USA}
\affiliation{Department of Physics, William \& Mary, P.O.~Box 8795
Williamsburg, Virginia 23187, USA}
%%%%%%%%%%
%
\date{\today}

%\pacs{
%12.38.-t %Quantum chromodynamics 
%11.15.Ha, % Lattice gauge theory
%12.38.Gc, % LQCD calculations
%12.38.-t, %Strong interactions in quantum chromodynamics
%13.15.+g, %Neutrino interactions with hadrons
%23.40.Bw, %Weak interactions in beta decay
%13.60.Fz %Elastic and Compton scattering 
%12.15.-y %Electroweak interactions 
% }
 
 %%%%%%%%%%%%%%%%%%%%%%%%%%%%%%%%%%%%
%	Preprint Number
%%%%%%%%%%%%%%%%%%%%%%%%%%%%%%%%%%%%
\preprint{JLAB-THY-21-3315}

\begin{abstract}
We present a model-independent framework to determine finite-volume corrections of matrix elements of spatially-separated current-current operators. We define these matrix elements in terms of Compton-like amplitudes, i.e.~amplitudes coupling single-particle states via two current insertions. We show that the infrared behavior of these matrix elements is dominated by the single-particle pole, which is approximated by the elastic form factors of the lowest-lying hadron. Therefore, given lattice data on the relevant elastic form factors, the finite-volume effects can be estimated non-perturbatively and without recourse to effective field theories.
For illustration purposes, we investigate the implications of the proposed formalism for a class of scalar theories in two and four dimensions. 
\end{abstract}
\maketitle
\nopagebreak
\section{Introduction \label{sec:intro}}
\noindent

In recent years, there has been significant progress in the direct determination of the structure and interactions of hadrons from quantum chromodynamics (QCD), the gauge theory of the strong nuclear force. This has been made possible through algorithmic and theoretical advances in lattice QCD (the discretization of QCD on a finite Euclidean hypercubic lattice), which is presently the only systematic \textit{ab initio} computational tool available to access hadronic properties. A key class of quantities that are at the cusp of being accessible using lattice QCD are generalized parton distributions (GPDs)~\cite{Ji:1996nm, Radyushkin:1996nd, Belitsky:2005qn}. GPDs capture the three-dimensional spatial distribution of hadronic constituents, and they generalize the elastic form factors that characterize the interactions of hadrons with electroweak probes. 

Our knowledge of GPDs, which are accessible through deeply-virtual
Compton scattering and deeply-virtual meson production, is restricted to certain kinematic regions  (for reviews see, for example, \cite{Favart:2015umi,dHose:2016mda,Kumericki:2016ehc}).
Until recently, lattice calculations were limited in their ability to determine GPDs, or their collinear counterparts, parton distribution functions (PDFs). The advent of new theoretical tools, including large-momentum effective theory (LaMET) \cite{Ji:2013dva}, factorizable matrix elements \cite{Ma:2014jla,Ma:2014jga,Ma:2017pxb}, and pseudodistributions \cite{Radyushkin:2016hsy,Radyushkin:2017cyf,Orginos:2017kos}, has offered, for the first time, the possibility of calculations of the three-dimensional structure of hadrons from the QCD Lagrangian. The common thread running through these formalisms is that structural information is obtained from matrix elements of currents that are separated in space, but local in time. This ensures that the matrix elements are insensitive to the time-signature of the correlation functions~\cite{Briceno:2017cpo}. The first lattice calculations of GPDs, applying the LaMET approach, appeared within the last year \cite{Alexandrou:2020zbe,Chen:2019lcm}. Recently, the first lattice calculations of generalized form factors, which correspond to the leading Mellin moments of GPDs, have appeared \cite{Alexandrou:2019ali,Bali:2018zgl}. For recent reviews see, for example, \cite{Monahan:2018euv,Zhao:2018fyu,Cichy:2018mum, Radyushkin:2019mye, Ji:2020ect}.

Lattice calculations of matrix elements relevant to GPDs present numerous technical challenges, including difficult signal-to-noise complications associated with fast-moving hadrons and significant systematic uncertainties. In addition to standard discretization errors, there are systematic uncertainties particular to PDFs and GPDs, including power divergences that arise from Wilson line operators on the lattice, higher twist effects, and enhanced finite volume effects. In Ref.~\cite{Briceno:2018lfj} we identified and a proposed a method for removing these enhanced finite volume effects in operators composed of spatially-separated currents. The first non-perturbative studies of finite volume effects have found mixed results \cite{Joo:2019bzr,Joo:2019jct,Alexandrou:2019lfo,Lin:2019ocg}, which may indicate that Wilson-line operators have reduced finite-volume effects relative to spatially-separated currents, first studied non-perturbatively in \cite{Sufian:2020vzb}.

Here we present a model-independent approach to determining the finite-volume error for matrix elements of spatially-extended two-current operators, relevant for the analysis of factorizable matrix elements. Following Ref.~\cite{Briceno:2019opb}, we define the matrix elements of spatially-separated currents in terms of Compton-like amplitudes, i.e.~amplitudes coupling single-particle states via two current insertions of, in principle, arbitrary Lorentz structure. Furthermore, we explain that ultimately one only needs the lowest-lying singularity of these amplitudes, namely the single-particle pole pieces. These pieces are completely constrained by the mass of the desired particle and the elastic form factors, which are among the quantities best constrained via lattice QCD, for example see Refs.~\cite{Gockeler:2003ay,Alexandrou:2006ru,Syritsyn:2009mx,Yamazaki:2009zq,Bratt:2010jn,Lin:2010fv,Collins:2011mk, Bhattacharya:2013ehc, Shanahan:2014cga,Shanahan:2014uka,Capitani:2015sba,Koponen:2017fvm,Chambers:2017tuf,Ishikawa:2018rew, Djukanovic:2019jtp,Jang:2019jkn,Park:2020axe, Alexandrou:2010hf,Rajan:2017lxk,Green:2017keo, Alexandrou:2017hac,  Shintani:2018ozy,  Jang:2019vkm, Shanahan:2018pib}. In other words, one can obtain the leading order finite-volume errors without making use of an effective field theory (EFT), which in general may have poor convergence. This framework is similar in spirit to that of Refs.~\cite{Hansen:2019rbh, Hansen:2020whp} for isolating the leading-order finite-volume error of the hadron-vacuum polarization contribution to the anomalous magnetic moment of the muon.

We find that, by comparing our results to the leading order effects determined using a scalar EFT, our current approach reinforces the conclusions of Ref.~\cite{Briceno:2018lfj}, without relying on a specific perturbative EFT analysis. We note, however, that the finite volume effects determined at leading order in the scalar EFT are generally of the same order of magnitude of, but parametrically smaller than, the effects determined from the full form factors. 

We start by presenting the main result and introducing the framework for spatially-separated current operators and Compton-like amplitudes in Sec.~\ref{sec:framework}. We apply our approach to a simple scalar model, testing several form factor parametrizations, in two dimensions in Sec.~\ref{sec:models2D} and in four dimensions in Sec.~\ref{sec:models4D}. We summarize in Sec.~\ref{sec:summary}.
 
\section{Framework \label{sec:framework}}

We begin by defining the infinite-volume matrix element \cite{Briceno:2018lfj}
\begin{equation}
\M_{\infty}(\pmb \xi, \textbf P_f, \textbf P_i)  \equiv \langle \textbf P_f \vert  \mathcal J(0, \pmb \xi) \mathcal J(0)    \vert \textbf P_i \rangle \,,
\label{eq:MIF}
\end{equation}
where $\vert \textbf P_i \rangle$ and $\vert \textbf P_f \rangle$ are the initial and final single-particle states, and $\xi^\mu =(0,{\pmb \xi})$ is the separation of the currents. For simplicity, we consider the case where the initial and final states are identical scalars of mass $m$, and we assume the currents, $\mathcal J$, to be scalars.  For further simplification, we set $ \pmb\xi=\xi\,\hat{z}$, where $\hat{z}$ is a unit vector in the $z$-direction. It is relatively straightforward to generalize these ideas to arbitrary Lorentz structures.

For clarity we quote here our main result. The difference between the infinite volume matrix element, Eq.~\eqref{eq:MIF}, and its finite-volume analogue is 
\begin{align}
\delta \M_{L}(\xi \hat{z}, \textbf P_f, \textbf P_i)  & \equiv \M_{L}(\xi \hat{z}, \textbf P_f, \textbf P_i)  - \M_{\infty}(\xi \hat{z}, \textbf P_f, \textbf P_i) 
\nn\\
&=
i
e^{-i P_{f,z}  \, \xi }
\int_q  e^{i q_z  (\xi -  L )}   
\,\frac{F(-(P_f-q)^2)
F(-(P_i-q)^2)
}{q^2-m^2+i\epsilon}\,
+\mathcal{O}(e^{-m L}),
\label{eq:master}
\end{align}
where we have introduced the $D$-dimensional measure, 
\begin{equation*}
\int_q = \int \frac{\dif^D q }{(2 \pi)^D}.
\end{equation*} 
Our result expresses the leading finite volume correction in terms of the scalar form factor $F(-(P_f-q)^2) = \langle \textbf P_f \vert   \mathcal J(0)    \vert \textbf q \rangle$. For arbitrary currents, $\mathcal{J}^A$ and $\mathcal{J}^B$, one can use Eq.~\eqref{eq:master} by replacing the scalar form factors with the corresponding elastic matrix elements, $\langle \textbf P_f \vert   \mathcal J^B(0)    \vert \textbf q \rangle \langle \textbf q \vert   \mathcal J^A(0)    \vert \textbf P_i \rangle$. Depending on the quantum numbers of the current, the $\mathcal J^A(0)    \vert \textbf P_i \rangle$ may not have the quantum numbers of the incoming state. In this  case, the pole appearing in the integrand will correspond to the mass of the lowest lying particle with the appropriate quantum numbers. 

In what follows, we show that this integral provides a model-independent determination of the coefficient of the leading-order finite volume correction for matrix elements of spatially separated currents. This correction is $\mathcal{O}(e^{-m |L-\xi|})$, as found using a scalar effective theory in Ref.~\cite{Briceno:2018lfj}.

To arrive at Eq.~\eqref{eq:master}, we follow Ref.~\cite{Briceno:2018lfj} and introduce the Fourier transform of this matrix element. In contrast to Ref.~\cite{Briceno:2018lfj}, however, we study this Fourier transform non-perturbatively using all-orders perturbation theory, rather than applying a perturbative analysis in the context of a scalar EFT. All-orders perturbation theory leads to results that are consistent with dispersive approaches and unitarity, and these enable us to introduce a model-independent definition of the leading finite-volume corrections to this matrix element.

\subsection{Derivation \label{sec:derive}}

\begin{figure}[t!]
\begin{center}
\includegraphics[width=\textwidth]{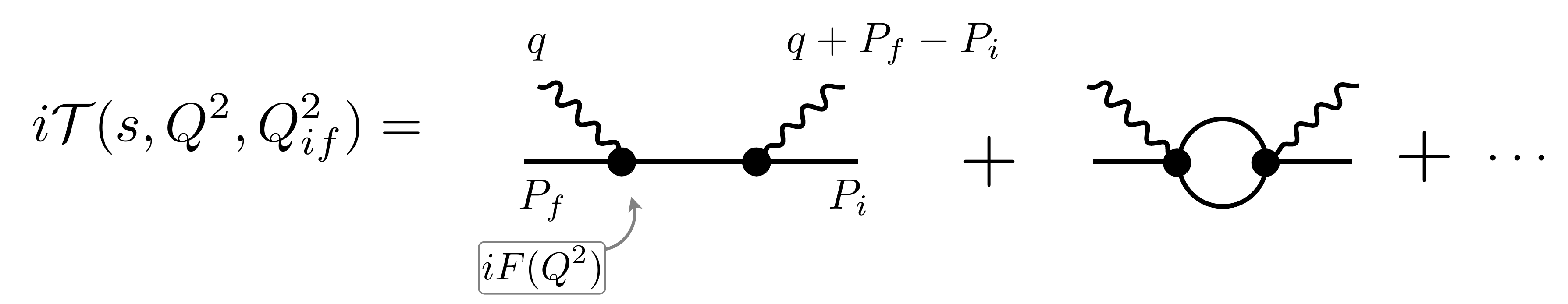}
\caption{Solid lines represent hadrons, the wiggly lines are external currents, the solid circles denote either elastic form factors or transition amplitudes.}
\label{fig:iT}
\end{center}
\end{figure}

In general, this matrix element can be defined as the Fourier transform of the ``Compton-like amplitude''\footnote{We reserve the term ``Compton amplitude'' for the specific case of the insertion of two vector currents.}, defined in Ref.~\cite{Briceno:2019opb},
 \begin{equation}
\label{eq:Minfgen}
\M_{\infty}(\pmb \xi, \textbf P_f, \textbf P_i) = \int_q e^{i \textbf q \cdot \pmb \xi} (-i) \T(s,Q^2,Q^2_{if} ) \,,
\end{equation}
where $Q^2=-q^2$ and $Q^2_{if} =-(P_f+q-P_i)^2$ are the virtualities of the two currents, and $s=(P_f+q)^2$. We define the Compton-like amplitude $\T$ to have factors of $i$ for each current insertion. In Fig.~\ref{fig:iT} we depict $\T$ diagrammatically, showing only the single-particle pole and the two particle contribution. Given that in general $\M_{\infty}$ is finite, the integral on the right hand side of Eq.~\eqref{eq:Minfgen} is convergent.

Working to all orders in perturbation theory, one can isolate the singularities of the amplitudes in the complex $s$-plane, as well as in the plane of the other kinematic variables. The closest singularity to the region of integration is the single-particle pole. As a result, this singularity describes the long-distance behavior of the matrix elements, which is the focus of this work. This contribution can be written in terms of the single-current matrix elements and the elastic hadronic form factors as
\begin{align}
i\T_{\rm pole}(s,Q^2,Q^2_{if} ) 
&=
i \langle \textbf P_f \vert   \mathcal J(0)    \vert \textbf q +\textbf P_f \rangle
\, \frac{i}{s-m^2+i\epsilon}
\, 
i \langle \textbf q +\textbf P_f \vert   \mathcal J(0)    \vert \textbf P_i \rangle
\nn\\
&=
 iF(Q^2)\,\frac{i}{s-m^2+i\epsilon}\,iF(Q^2_{if}),
\label{eq:Tpole}
\end{align}
which we depict in Fig.~\ref{fig:iT}.

%%%%%%%%%%%%%%%%%%%
%%%%%%%%%%%%%%%%%%%
%%%%%%%%%%%%%%%%%%%
%%%%%%%%%%%%%%%%%%%

Having expressed the matrix element in terms of the Compton-like amplitude, we can now evaluate the finite-volume corrections. There are two classes of finite-volume effects. The first are standard errors associated with virtual particles ``\emph{wrapping around the volume}''. In Appendix~\ref{appendixA} we show that these effects can be encoded by replacing $\T$ with its finite-volume analogue $\T_L$ introduced in Ref.~\cite{Briceno:2019opb}. These effects lead to the standard $\mathcal{O}(e^{-mL})$ errors. The second class of finite-volume artifacts arise from the spatial separation of the two currents, $\boldsymbol{\xi}$. As explained in Ref.~\cite{Briceno:2018lfj}, this class of errors scale as $\mathcal{O}(e^{-m|L-\xi|})$. Therefore, for $mL\gg 1$ the latter will be the dominant error. Here we explain how to determine the prefactor multiplying the $e^{-m|L-\xi|}$ behavior, without recourse to an effective field theory.

The finite-volume analogue of Eq.~\eqref{eq:Minfgen} can be written as the finite-volume Fourier transform of $\T_L$. As a result, we can write the finite-volume correction to the matrix element as
\begin{align}
\delta \M_{L}(\pmb \xi, \textbf P_f, \textbf P_i)  
& =
\frac{1}{L^{D-1}}\sum_{\textbf q} \int \frac{\dif q^0 }{2 \pi} e^{i \textbf q \cdot \pmb \xi} (-i) \T_L(s,Q^2,Q^2_{if} )
-\int_q e^{i \textbf q \cdot \pmb \xi} (-i) \T(s,Q^2,Q^2_{if} )
%%%%%%%%%%%%%%%%%%%%%%%%%%%%%%%%
\nn\\
& =
\frac{1}{L^{D-1}}\sum_{\textbf q} \int \frac{\dif q^0 }{2 \pi} e^{i \textbf q \cdot \pmb \xi} (-i) \T(s,Q^2,Q^2_{if} )
-\int_q e^{i \textbf q \cdot \pmb \xi} (-i) \T(s,Q^2,Q^2_{if} )
+\mathcal{O}(e^{-m L})
%%%%%%%%%%%%%%%%%%%%%%%%%%%%%%%%
\nn\\
& =
\sum_{ \textbf n\neq 0 } \int_q  e^{i \textbf q \cdot (\pmb \xi +  L \textbf n)}   
\,(-i) \T(s,Q^2,Q^2_{if} )
+\mathcal{O}(e^{-m L})
\label{eq:dML_gen}
\\
& =
\sum_{ \textbf n\neq 0 } 
\M_{\infty}(\pmb \xi +  L \textbf n, \textbf P_f, \textbf P_i)
+\mathcal{O}(e^{-m L})
.
\label{eq:masterv2}
\end{align}
where in the second equality we made use of the arguments presented in Appendix~\ref{appendixA} to replace $\T_L$ with $\T$ up to $\mathcal{O}(e^{-m L})$ errors. We have used the Poisson summation formula in the third equality. 
The last equality, although formally correct, is not in general useful. This result states that the finite-volume corrections of the matrix elements at $\pmb \xi$ depend on the value of ${\cal M}(\pmb \xi +  L \textbf n,\textbf P)$ where $|\textbf n|\neq0$, which is a-priori unknown. Therefore, we use instead the second-to-last equality, Eq.~\eqref{eq:dML_gen}, to estimate the large distance value of the matrix element and, from this, infer the finite-volume effects.

Depending on the specific choice of $\pmb \xi$, the sum over $\textbf n$ includes finite-volume errors that are $\mathcal{O}(e^{-m L})$ or smaller, which we neglect in this analysis. In the following derivation, we choose $\pmb \xi=\xi\hat{z}$, which is the case most typically used in calculations. With this choice, the finite-volume correction reduces to 
\begin{align}
\delta \M_{L}( \xi \hat{z}, \textbf P_f, \textbf P_i)  
& =
\int_q  e^{i \textbf q \cdot\hat{z}  ( \xi -  L )}   
\,(-i) \T(s,Q^2,Q^2_{if} )
+\mathcal{O}(e^{-m L})
\label{eq:dML_gen2}
%%%%%%%%%%%%%%%%%%%%%%%%%%%%%%%%
\\
& =
\M_{\infty}( \xi -  L , \textbf P_f, \textbf P_i)
+\mathcal{O}(e^{-m L}).
\end{align}
The generalization to other geometries is straightforward, and requires including other modes in the finite volume sum. For example, if $\pmb \xi=\frac{\xi}{\sqrt{2}} (1,1,0)$, then $\textbf n=(-1,0,0)$ and $\textbf n=(0,-1,0)$ both contribute equally. For moderate values of $\pmb \xi$, i.e.  $\xi < L/2$, these two modes would provide the leading order contributions to the finite-volume errors.

Having established the relationship between the desired matrix elements and the Compton-like amplitude, we can identify the relation between the long-range contributions to the matrix elements and the low-energy contributions to the amplitude. In particular, one expects the integral to be saturated by the small $|\textbf{q}|$ region. Thus, although the pole contribution is not, in general, a good description of $\M_{\infty}$ for small values of $\xi$, this contribution does provide a reasonable approximation of the finite-volume corrections to $\M_{\infty}$. In other words, we approximate Eq.~\eqref{eq:dML_gen2} using Eq.~\eqref{eq:Tpole} for the pole contribution,
\begin{align}
\delta \M_{L}(\xi \hat{z}, \textbf P_f, \textbf P_i)  
& = \int_q  e^{i \textbf q \cdot\hat{z}  (\xi -  L)}   
(-i)\ \,\frac{F(Q^2)\,F(Q^2_{if})}{-(q+P_f)^2+m^2-i\epsilon}\,
+\mathcal{O}(e^{-m L}),
\label{eq:dML_F}
\end{align}
which is our main result, Eq.~\eqref{eq:master}.

This approximation can be further justified as follows. The expected form of this integral, Eq.~\eqref{eq:dML_F}, is $\mathcal{O}(e^{-m |L-\xi|})$, as was shown in Ref.~\cite{Briceno:2018lfj}. The neglected terms in $\T$ will have singularities associated with a higher energy scale, $M\gg m$. In general this energy scale could correspond to excited states or thresholds. Performing a spectral decomposition for these singularities, one can immediately conclude that these contributions will lead to finite-volume corrections of the form $\mathcal{O}(e^{-M |L-\xi|})$. For moderately small values of $\xi$, these errors can be safely neglected.

Expressing the finite volume corrections in this form has several advantages over the perturbative EFT expansion applied in Ref.~\cite{Briceno:2018lfj}. First and foremost, this representation is non-perturbative in the dynamics. We have made no assumption about the power counting of any underlying EFT. Instead, we have made the purely kinematic assumption that the single-particle pole is the dominant singularity. Moreover, this can be systematically improved by including further singularities, starting with the two-particle cut, the form of which was recently derived in Ref.~\cite{Briceno:2019opb}. Finally, this representation extends straightforwardly to finite-volume effects for matrix elements involving initial and final states that have different momenta, which is particularly relevant for calculations of GPDs.

In practice, the main challenge to evaluating finite-volume corrections using this representation is that, formally, knowledge of the form factors over a large kinematic region is required. In fact, each integral in Eq.~\eqref{eq:dML_F} ranges from negative to positive infinity. State-of-the-art lattice calculations generally span a small region at low momentum transfer, but calculations up to $Q^2 \simeq 6.0\;\mathrm{GeV}^2$ have been carried out \cite{Lin:2010fv, Chambers:2017tuf, Koponen:2017fvm}. As a result, one might worry that this framework could be unrealistic to implement. But, as has been mentioned before, these integrals are dominated by the small $\textbf q$ region. Thus, we envision using parametrizations of the form factors that describe the lattice QCD form factors accurately and vanish rapidly enough as $|\textbf q|\to \infty$. In the following sections we test these ideas for a simple dipole parametrization of the form factors.

Having a covariant parametrization for the form factor, one may proceed to evaluate analytically the integral shown in Eq.~\eqref{eq:dML_F} using the techniques used in Ref.~\cite{Briceno:2018lfj}. Alternatively, one can further approximate Eq.~\eqref{eq:dML_F} by evaluating the $q_0$ integral and only keeping the pion pole contribution, to give
\begin{align}
\delta \M_{L,\mathrm{pole}}(\xi \hat{z}, \textbf P_f, \textbf P_i)  
& \equiv
 \int \frac{\dif^{D-1} {\textbf q} }{(2 \pi)^{D-1}}  e^{i \textbf q \cdot\hat{z}  (\xi -  L)}   
\frac{F(Q^2)\,F(Q^2_{if})}{2\sqrt{(\textbf q+\textbf P_f)^2+m^2}} 
\bigg|_{q_0=-E+\sqrt{(\textbf{q}+\textbf{P}_f)^2+m^2}},
\label{eq:dML_F_pi}
\end{align}
where the subscript ``$\mathrm{pole}$'' indicates that only the pion pole contribution has been retained. 

\subsection{Comparison with scalar EFT}

Before we apply this formalism to a scalar model, we note that we can directly compare the form-factor representation of Eq.~\eqref{eq:dML_F} to the corresponding expression obtained from a scalar EFT at leading order studied in Ref.~\cite{Briceno:2018lfj}. To compare these results, we replace each form factor in the Compton-like amplitude, Eq.~\eqref{eq:Tpole}, by its value evaluated at $Q^2 = 0$, i.e.~with the charge of the particle $iF(0) = g$, and set the initial and final momenta equal, $P=P_i=P_f$. With these replacements, the Compton-like amplitude reduces to
\begin{align}
\T_{\rm LO}(s,Q^2,Q^2 ) = -g\,\frac{1}{s-m^2+i\epsilon}\,g.
\end{align}
Inserting this into Eq.~\eqref{eq:dML_F}, we find
 \begin{align}
\label{eq:ML_L0}
\delta\M_{L;\rm LO}(\xi \hat{z}, \textbf P) 
&=   \int_q e^{i \textbf q \cdot\hat{z}  (\xi -  L)}    \,
(-i)\frac{g^2}{-(q+P)^2+m^2-i\epsilon}\, \,,\nn\\
&=  
\int_{q_E} e^{i \textbf q \cdot\hat{z} (\xi -  L)}     \,\frac{g^2}{(q_E+P_E)^2+m^2}\, \,,
\end{align}
where we have replaced $q=iq_E$, $P=iP_E$. This is the same expression presented in Eq.~(15) of Ref.~\cite{Briceno:2018lfj} for the leading order contribution. 

In the following sections, we refer to this prescription as the ``charge prescription'' and compare the results for the corresponding finite volume effects with those estimated using  parametrizations of the form factors.

\section{Model calculations in 2D \label{sec:models2D}}

The form factor representation of the finite volume corrections, Eq.~\eqref{eq:dML_F}, is model independent and depends only on the kinematic approximation that the integral of the Compton-like amplitude is dominated by the closest pole singularity. In this section we test the implications of this representation in a scalar theory in two dimensions. For simplicity, we set the initial and final momenta to be equal, and use the notation $\textbf{P}=\textbf{P}_i=\textbf{P}_f = P_z\hat{z} $, which is the case relevant to calculations of collinear hadron structure, and label matrix elements by $P_z$. The extension to the off-forward case, relevant to calculations of GPDs, is straightforward. 

Based on the results of Ref.~\cite{Briceno:2018lfj}, this formalism is likely to be most immediately useful for the estimation of the finite-volume effects for matrix elements of the pion. With this in mind, although the following discussion is quite general and assumes only that the hadronic state has zero spin, we will refer to this state as the pion.  

Assuming the single-particle pole dominates the integral, in two dimensions the infinite-volume matrix element in Eq.~\eqref{eq:Minfgen} can be approximated as
\begin{equation}
\M^{2D}_{\infty}( \xi\,\hat{z}, P_z\hat{z}) 
\approx  i \int\frac{\dif^2q}{(2\pi)^2} e^{iq_z\xi}  \,
\frac{\left(F(Q^2)\right)^2 }{(q+P)^2-m^2+i\epsilon} .
\label{eq:dMI_F2D}
\end{equation}
Similarly, the finite-volume matrix element, given in Eq.~\eqref{eq:dML_F}, reduces to a two dimensional integral, 
\begin{align}
\delta \M_{L}(\xi\,\hat{z}, P_z\hat{z})  
& \approx
\int \frac{\dif^2 q }{(2 \pi)^2} e^{i q_z (\xi -  L)}   
\ 
\frac{\left(F(Q^2)\right)^2 }{(q+P)^2-m^2+i\epsilon} .
\label{eq:dML_F2D}
\end{align}

In order to explore the implications for this formalism, we evaluate the integral of Eq.~\eqref{eq:dML_F2D} using three different approximations. First, we parametrize the form factor using a dipole form and evaluate these integrals exactly. Second, using the dipole form, we approximate the integral even further by evaluating only the pion pole contribution. Finally, we compare our results to the ``charge prescription''.

\subsection{Dipole form factor \label{ssec:2d}}

One standard parametrization for form factors is the dipole form,
\begin{equation}
iF(Q^2) = -ig\frac{m_F^2}{q^2-m_F^2+i\epsilon},
\label{eq:dipole}
\end{equation}
where $m_F$ is typically attributed to a resonance with the quantum numbers of the current. Such a pole, in general, violates unitarity. Instead, unless the form factor has the quantum numbers of a single particle states [e.g.~the axial vector in QCD], the closest singularity of the form factor will be the first branch-cut associated with multi-particle production in the time-like region. Despite this, the dipole approximation provides a reasonable prescription of form factors in the spacelike region. 

With this expression for the form factor, the integral of Eq.~\eqref{eq:dML_F2D} can be expressed as
\begin{equation}
\M^{2D}_{\infty,\rm dip.}(\xi\,\hat{z}, P_z\hat{z}) =ig^2m_F^4\int\frac{\dif^2q}{(2\pi)^2}  \frac{e^{i q_z\xi}}{(q^2-m_F^2+i\epsilon)^2((q+P)^2-m^2+i\epsilon)}.
\label{eq:dip_2D}
\end{equation}

This integral can be evaluated using standard Schwinger tricks, as carried out in, for example, Ref.~\cite{Briceno:2018lfj}. Instead, we choose to first evaluate the $q^0$ integral using Cauchy's residue theorem. The integrand has four poles, 
\begin{align}
q^0&=\pm(\omega_{qF}-i\epsilon),
\hspace{3cm}
q^0=-E\pm(\omega_{qP}-i\epsilon),
\end{align}
where we have introduced $E=\sqrt{P_z^2+m^2}$ as the energy of the pion and defined 
\begin{align}
\omega_{qF} =& \sqrt{q_z^2+m_F^2},
\hspace{3cm}
\omega_{qP}  =  \sqrt{(q_z+P_z)^2+m^2}.
\end{align}
The first of these poles arises from the form factor, and the second from the pion propagator.

Closing the contour in the upper half plane, we obtain
\begin{equation}
\label{eq:I2}
\M^{2D}_{\infty, \rm dip.}(\xi\,\hat{z}, P_z\hat{z}) 
= 
\M^{2D}_{\infty,\rm pole}(\xi\,\hat{z}, P_z\hat{z})+\M^{2D}_{\infty,\rm FF}(\xi\,\hat{z}, P_z\hat{z})
\end{equation}
where $\M^{2D}_{\infty,\rm pole}$ and $\M^{2D}_{\infty,\rm FF}$ are the pion and form factor pole contributions, respectively, and are explicitly defined as
\begin{align}
\M^{2D}_{\infty,\rm pole}(\xi\,\hat{z}, P_z\hat{z})
&=
\frac{(gm_F^2)^2}{2\pi}
\int_{-\infty}^{\infty}\dif q_z\,e^{iq_z \xi}\frac{1}{2\omega_{qP}}\frac{1}{\big((E-\omega_{qP})^2- \omega_{qF}^2\big)^2}
,
\label{eq:M2pi}\\
%%%%%%%%%%%%%%%%%%%%%%%%
\M^{2D}_{\infty,\rm FF}(\xi\,\hat{z}, P_z\hat{z})&=
\frac{(gm_F^2)^2}{2\pi}
\int_{-\infty}^{\infty}\dif q_z\,e^{iq_z \xi}
\frac{1}{4\omega_{qF}^3}
\left[
\frac{\omega_{qP}^2 -  (E +\omega_{qF})\, (E +3\,\omega_{qF})}
{
\big((E + \omega_{qF})^2-\omega_{qP}^2\big)^2
}
\right].
\label{eq:M2FF}
\end{align}
Note that Eq.~\eqref{eq:M2pi} is just the two-dimensional and model-dependent version of Eq.~\eqref{eq:dML_F_pi}. We separate the pole contributions to the matrix element so that we can see the role of each term, and in particularly showcase the dominance of the pion pole contribution. One can then evaluate the remaining integrals numerically.

In a finite volume the momentum integral is replaced a sum of lattice modes. A similar analysis to the infinite-volume case leads to an expression for the finite volume corrections,
\begin{equation}
\label{eq:I2FV2D}
\delta \M^{2D}_{L,\rm dip.}(\xi\,\hat{z}, P_z\hat{z}) = \delta \M^{2D}_{L,\mathrm{pole}}(\xi\,\hat{z}, P_z\hat{z})+\delta \M^{2D}_{L,\rm FF}(\xi\,\hat{z}, P_z\hat{z})
\end{equation}
where, 
\begin{align}
\delta \M^{2D}_{L,\mathrm{pole}}(\xi\,\hat{z}, P_z\hat{z})
&=
\frac{(gm_F^2)^2}{2\pi}
\int_{-\infty}^{\infty}\dif q_z\,
\frac{e^{i q_z (\xi -  L)}    }{2\omega_{qP}}\frac{1}{\big((E-\omega_{qP})^2- \omega_{qF}^2\big)^2}
,
\label{eq:M2piFV}\\
%%%%%%%%%%%%%%%%%%%%%%%%
\delta \M^{2D}_{L,\rm FF}(\xi\hat{z}, P_z\hat{z})&=
\frac{(gm_F^2)^2}{2\pi}\int_{-\infty}^{\infty}\dif q_z\,
\frac{e^{i q_z (\xi -  L)}   }{4\omega_{qF}^3}
\left[
\frac{\omega_{qP}^2 -  (E +\omega_{qF})\, (E +3\,\omega_{qF})}
{
\big((E + \omega_{qF})^2-\omega_{qP}^2\big)^2
}
\right].
\label{eq:M2FFV}
\end{align}
Note that, consistent with our main results, Eq.~\eqref{eq:master}, we have only kept the finite-volume mode that leads to the leading finite-volume error. All other modes give errors of $\mathcal{O}(e^{-mL})$ or smaller for moderately small values of $\xi$, and are therefore neglected.

\subsection{Charge prescription \label{ssec:charge}}

As a comparison, we also evaluate the charge prescription, considered in Ref.~\cite{Briceno:2018lfj}. From Eq.~\eqref{eq:ML_L0}, we can write this in two dimensions as
\begin{align}
\delta \M^{2D}_{L,\rm LO}(\xi \hat{z}, P_z\hat{z})   
 =g^2
\int \frac{\dif^2 q_E }{(2 \pi)^2}
e^{i q_z (\xi -  L)} 
\frac{1 }{(q_E+P_E)^2+m^2},
\label{eq:ML_L0_2D}
\end{align}
Introducing a Schwinger representation of the propagator, and carrying out the momentum and Schwinger integrals, leads to
\begin{align}
\delta \M^{2D}_{L,\rm LO}(\xi \hat{z}, P_z\hat{z})   = \frac{g^2}{2\pi}e^{-i P_z( \xi -  L )}K_0(m|\xi -L |),
\label{eq:M2LO_P}
\end{align}
where $K_n(z)$ is the modified Bessel function of the second kind.

\begin{figure}[t!]
\begin{center}
\includegraphics[width=\textwidth]{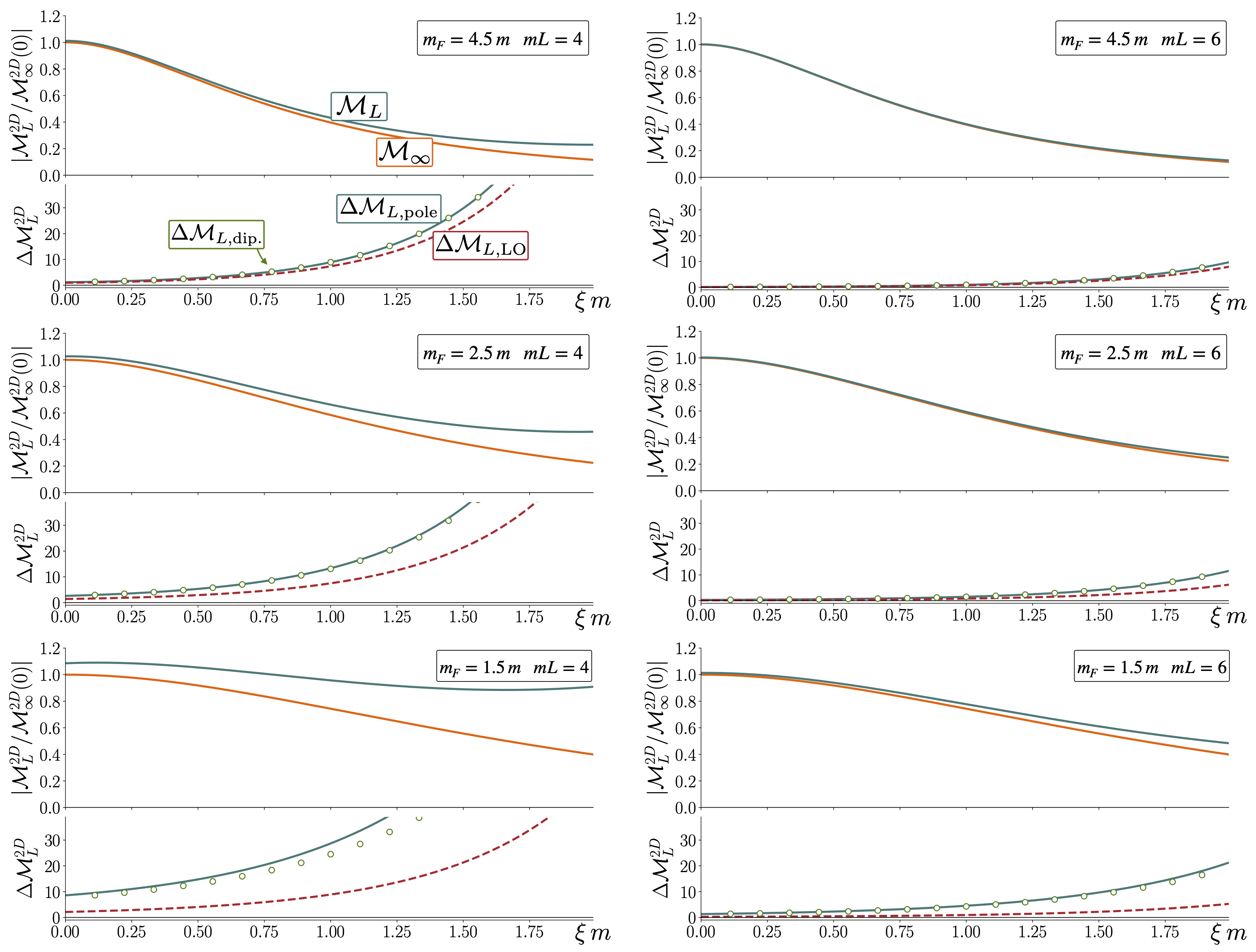}
\caption{The two-dimensional finite-volume matrix elements versus the infinite-volume matrix element as a function of the separation of the two currents for a range of parameters, with initial and final momentum fixed to zero. For each set of parameters, the bottom half of the panels show the percentage finite-volume errors as defined in Eq.~\eqref{eq:DeltaM}. }
\label{fig:pole_L_dep}
\end{center}
\end{figure}

\begin{figure}[t!]
\begin{center}
\includegraphics[width=\textwidth]{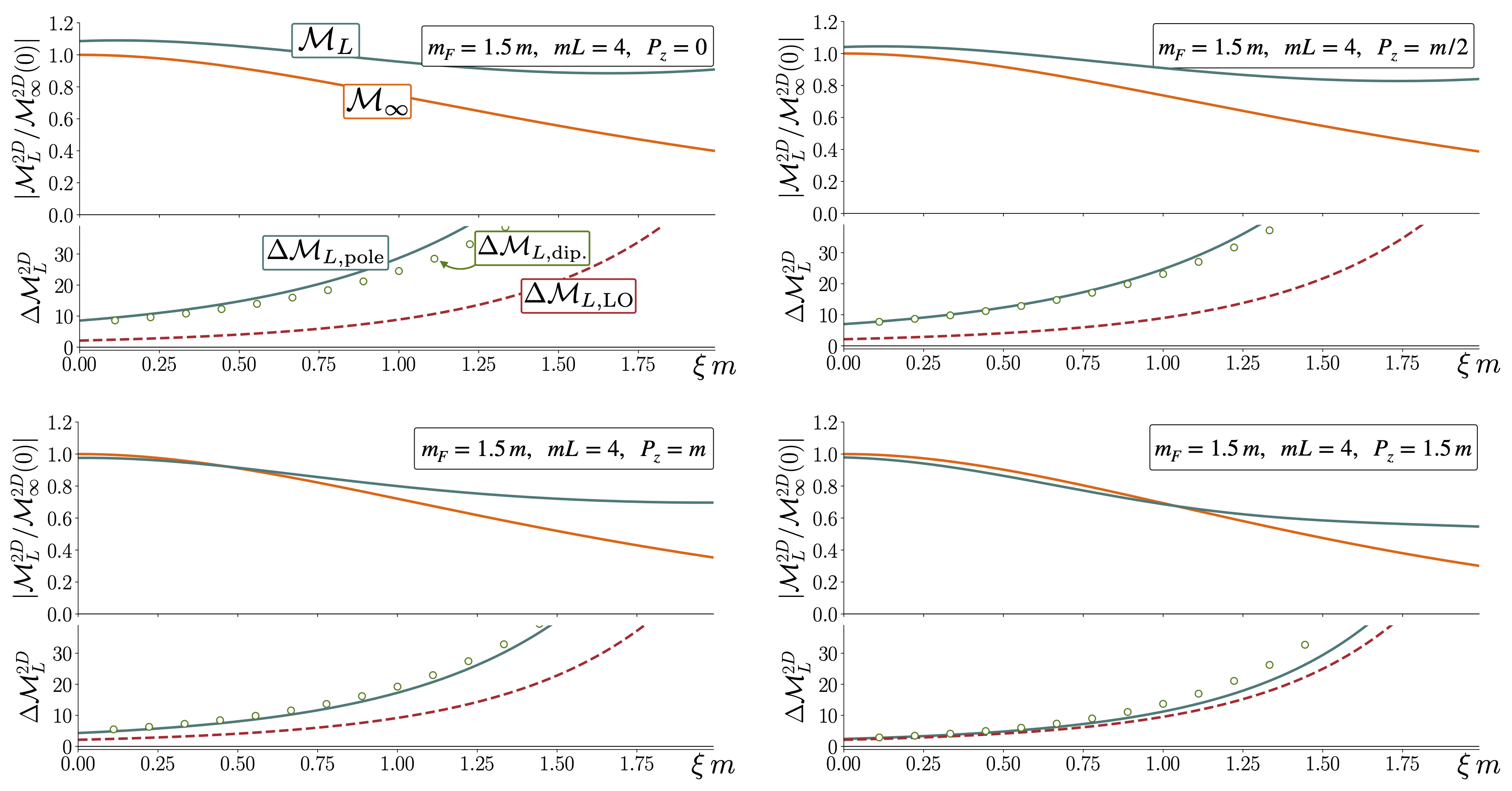}
\caption{Same as in Figure~\ref{fig:pole_L_dep}, but the parameters $m_F$ and $L$ are fixed to be $1.5m$ and $4/m$ respectively. Instead, the momenta are varied from $0$ to $1.5\,m$.}
\label{fig:mom_dep}
\end{center}
\end{figure}

\subsection{Results in two dimensions \label{sec:2dresults}}

We have presented three schemes for estimating the finite-volume error. These correspond to using the covariant dipole form factor, keeping only the pion pole, and the charge prescriptions, defined by Eqs.~\eqref{eq:I2FV2D}, \eqref{eq:M2piFV}, and \eqref{eq:M2LO_P}, respectively.
We compare our results for the three schemes in Figures \ref{fig:pole_L_dep}, and \ref{fig:mom_dep}. In Figure \ref{fig:pole_L_dep} we compare the finite-volume and infinite-volume matrix elements as functions of the separation of the two currents, with initial and final momentum fixed to zero. We use six sets of parameters $\{m_F, L, z\}$, expressed each combination in terms of $m$. We state our choices for the combination of parameters in the upper right corner of each panel. 

For each choice of parameters we calculate $\delta \M^{2D}_{L, \rm dip.}$, $\delta\M^{2D}_{L}$, and $\delta\M^{2D}_{L}$. For the finite-volume matrix element we use $\delta \M^{2D}_{L, \rm dip.}$. Both $\M^{2D}_{\infty, \rm dip.}$ and $\M^{2D}_{L, \rm dip.}$ are shown in units of $\M^{2D}_{\infty, \rm dip.}$ evaluated at $\xi=0$. We highlight that at small current separations, $\xi\sim 0$, the leading finite volume correction is $e^{-|L-\xi| m}\sim e^{-Lm}$. In other words, for small $\xi$ the dominant error is of the order of the suppressed errors. For clarity of presentation, however, we only keep the error that scales $e^{-|L-\xi| m}$ in the plots, even when $\xi= 0$.

We determine the percentage finite-volume error
\begin{align}
\Delta \M^{2D}_L \equiv \left| \frac{\delta \M^{2D}_{L} }{\M^{2D}_{\infty} } \right|\times 100,
\label{eq:DeltaM}
\end{align}
for each choice of parameters and framework for evaluating the finite-volume corrections. 
We plot this percentage as a function of $\xi$ in the bottom half of each one of the panels.

From Figure \ref{fig:pole_L_dep} we observe two important features. First, regardless of the choice of the form factor, the charge prescription consistently underestimates the finite-volume corrections. Second, as expected, the pion pole prescription dominates the finite-volume corrections. This is because, as explained in Sec.~\eqref{sec:derive}, the pion pole is the closest singularity of the Compton amplitude to the region of integration. As noted in Ref.~\cite{Briceno:2018lfj}, for the choice of $m L \sim 4$, finite-volume effects can be sizable. Here we find that for a form factor with a small pole mass, i.e. $m_F < 2.5 m$, these effects can be of the order of 10\%-20\% even for small values of $\xi$, $\xi m \simeq 0.5$. These effects are reduced to the percent level for a slightly larger volume of $m L\sim 6$. 

These larger finite-volume effects at smaller values of $m_F$ can be understood in terms of the singularities of the integrand. The smaller the value of $m_F$, the closer the pole of the form factor is to the kinematic region of integration in Eq.~\eqref{eq:M2pi}. 

In Figure \ref{fig:mom_dep} we explore the momentum dependence of matrix elements with $m_F=1.5 m$ and $mL=4$, which suffer the largest finite-volume effects. Once again, the plots illustrate the charge prescription consistently underestimates of the leading finite volume effects, relative to the form factor models. Furthermore, one observes that finite-volume errors are further suppressed at increasingly large momenta.

\section{Model calculations in four dimensions \label{sec:models4D}}
Here we follow similar steps to those presented in Sec.~\ref{sec:models2D} for the four-dimensional case. We again set the initial and final momenta to be equal and align it along the $\hat{z}$ axis, $\textbf{P}=\textbf{P}_i=\textbf{P}_f = P_z\hat{z} $. As before, we also align the current displace along the $\hat{z}$ axis, $\pmb \xi = \xi \hat{z}$. 
Applying the same approximations discussed before, the four-dimensional infinite-volume matrix element in Eq.~\eqref{eq:Minfgen} can be written as
\begin{equation}
\M_{\infty}(\xi\hat{z}, P_z\hat{z}) 
\approx  i \int\frac{\dif^4q}{(2\pi)^4} e^{iq_z \xi}  \,
\frac{\left(F(Q^2)\right)^2 }{(q+P)^2-m^2+i\epsilon} .
\label{eq:dMI_F4D}
\end{equation}
Similarly, the finite-volume counterpart, Eq.~\eqref{eq:dML_F}, is equal to
\begin{align}
\delta \M_{L}(\xi\hat{z}, P_z\hat{z})  
& \approx
i\int \frac{\dif^4 q }{(2 \pi)^4}  e^{i q_z (\xi- L)}   
\ 
\frac{\left(F(Q^2)\right)^2 }{(q+P)^2-m^2+i\epsilon} .
\label{eq:dML_F4D}
\end{align}

Using the dipole form factor defined in Eq.~\eqref{eq:dipole}, the infinite-volume matrix element can be written as
\begin{equation}
\M_{\infty,\rm dip.}(\xi\hat{z}, P_z\hat{z}) =ig^2m_F^4\int\frac{\dif^4q}{(2\pi)^4}  \frac{e^{iq_z \xi}}{(q^2-m_F^2+i\epsilon)^2((q+P)^2-m^2+i\epsilon)}.
\label{eq:dip_4D}
\end{equation}
In principle, one could evaluate the $q_0$ integral and obtain two contributions, which would be the three-dimensional analogues of Eqs.~\eqref{eq:M2piFV} and \eqref{eq:I2FV2D}. It is more convenient, however, to retain the covariant four-dimensional integral, introduce a Feynman parameter followed by the Schwinger parameterization, as in Ref.~\cite{Briceno:2018lfj}. This allows one to reduce the four-dimensional integral to a single integral over the Feynman parameter. Using Eqs.~(16), (A9), and (A11) in Ref.~\cite{Briceno:2018lfj}, one can show that this is equal to 
\begin{align}
\M_{\infty,\rm dip.}(\xi\hat{z}, P_z\hat{z}) =\xi\frac{(gm_F^2)^2}{(4\pi)^4}\int_0^1dx
\frac{x\, e^{i (x-1) P_z\xi}}{\sqrt{xm_F^2+m^2(1-x)^2}}
K_1\left(\xi\,\sqrt{xm_F^2+m^2(1-x)^2}\right).
\label{eq:dip_4D_v2}
\end{align}
Then, from the last line of Eq.~\eqref{eq:dML_gen}, one can write the finite-volume correction in terms of the sum over the mirror-image contributions of this integral, with the largest contribution coming from the nearest image,
\begin{align}
\M_{L,\rm dip.}(\xi\hat{z}, P_z\hat{z}) & =
\M_{\infty,\rm dip.}(\xi - L, P_z\hat{z}).
\end{align}

We can compare this with the charge prescription in four dimensions, Eq.~\eqref{eq:ML_L0}, which was considered in Ref.~\cite{Briceno:2018lfj}. Applying Eq.~(16) of Ref.~\cite{Briceno:2018lfj} one readily obtains
  \begin{align}
\label{eq:ML_L0_4D}
\delta\M_{L;\rm LO}(\xi\hat{z}, P_z\hat{z}) 
&=\int \frac{\dif^4 q_E }{(2 \pi)^4} e^{i q_z  (\xi -  L)}    \,\frac{g^2}{(P_E+q_E)^2+m^2}\, \,,
\nn\\
& = \frac{m\,g^2}{4\pi^2}
\,e^{-i \textbf{P}\cdot (\xi\,\hat{z} +  L \textbf n)}
\frac{K_1\left(|\xi- L |\,m\right)}{|\xi-  L|}.
\end{align}
This can be further simplified by using the fact that $e^{i|\textbf{P}|  L} =e^{i 2\pi |\textbf{m}|}=1$,
\begin{align}
\label{eq:ML_L0_4D_v2}
\delta\M_{L;\rm LO}( \xi\hat{z}, P_z\hat{z}) 
& = \frac{m\,g^2}{4\pi^2}
\,e^{-i P_z  \xi }
\frac{K_1\left(| L -\xi |\,m\right)}{| L -\xi  |} .
\end{align}

\begin{figure}[t!]
\begin{center}
\includegraphics[width=1\textwidth]{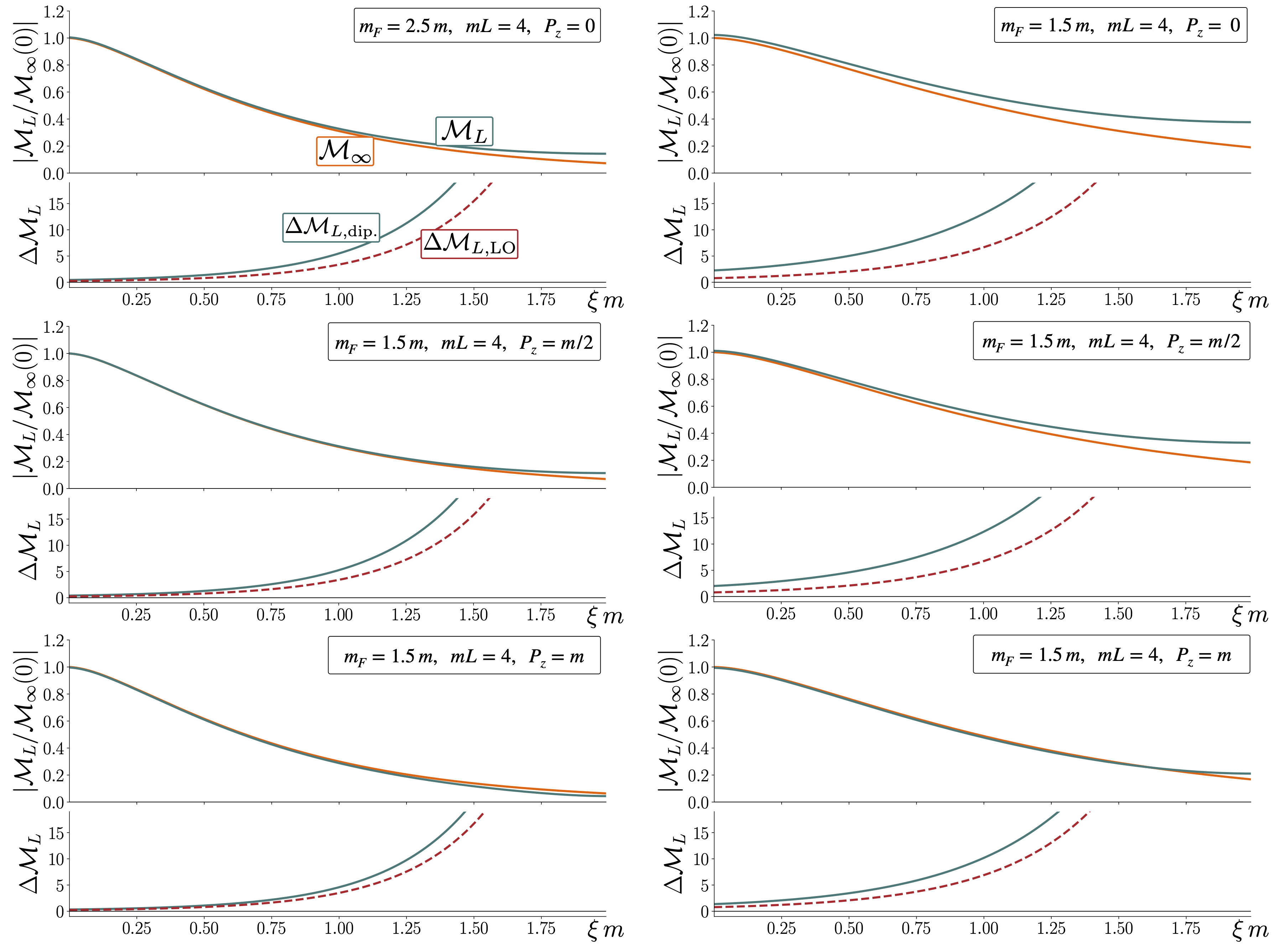}
\caption{We plot the four-dimensional finite-volume matrix element versus the infinite-volume one as a function of the separation of the two currents for a range of momenta, with all other parameters fixed. }
\label{fig:threeD_L_mom_dep}
\end{center}
\end{figure}

We now have the ingredients needed to compare numerical results. Following Sec.~\ref{sec:2dresults}, we expect the largest finite-volume artifacts at smaller values of $m_F$ and therefore focus our attention on $m_F=2.5\,m$ and $1.5\,m$. In Figure~\ref{fig:threeD_L_mom_dep} we show results for these two values of $m_F$ and a range of external momenta. As in the two-dimensional case, we see that the leading-order contribution underestimates the finite-volume artifacts. However, the results show  that the leading-order contribution captures the approximate error much more effectively in four dimensions than it does in two dimensions. Furthermore, the difference between the leading-order contribution and the full dipole form vanishes relatively quickly for increasing values of the external momenta. 

Most importantly, the overall magnitude of the error is about four times smaller in four dimensions than in two dimensions. This can be seen by comparing the values of these two cases for $m_F=1.5\,m$, $\xi=0$ and $P_z=0$ in Figures~\ref{fig:pole_L_dep} and \ref{fig:threeD_L_mom_dep}. One can reconcile this observation by again considering the singularities of the integrands. The integrands in the two-dimensional and four-dimensional integrals, Eqs.~\eqref{eq:dip_2D} and ~\eqref{eq:dip_4D} respectively, are identical; the only difference is the dimension of the integral. The two additional angular integrals in the four-dimensional integral soften the singularity of the integrand and consequently the magnitude of the finite volume corrections.

\section{Conclusions\label{sec:summary}}

Recent theoretical developments have enabled the determination of hadron structure directly from QCD. Calculations of collinear hadron structure have matured sufficiently that a detailed understanding of systematic uncertainties is necessary to compare these calculations to experimental data. In Ref.~\cite{Briceno:2018lfj} we suggested that the spatially extended composite operators used in many lattice calculations of hadron structure may induce enhanced finite volume effects. Here we developed an EFT-independent framework for estimating those finite volume effects.

By defining the matrix elements of spatially-separated currents in terms of a Compton-like amplitude, introduced in Ref.~\cite{Briceno:2019opb}, we argued that the infrared behavior of these matrix elements is dominated by the single-particle pole. This contribution can be determined from the elastic form factors of the lowest-lying hadronic state with the appropriate quantum numbers. This provides an opportunity to estimate finite volume effects without relying on an underlying EFT that may have, in general, poor convergence. 

We studied this methodology in simple scalar models, in two and four dimensions, focusing on the case of two spatially-separated scalar currents. We found that, by comparing our results to those derived from a scalar EFT in Ref.~\cite{Briceno:2018lfj}, our current approach reinforces the conclusions of Ref.~\cite{Briceno:2018lfj}. We note, however, that the finite volume effects determined at leading order in the scalar EFT are generally smaller than determinations that incorporate form factors.

\section{Acknowledgements}
The authors would like to thank M. T. Hansen and J. V. Guerrero for useful discussions. 
RAB and CJM are supported in part by USDOE grant No. DE-AC05-06OR23177, 
under which Jefferson Science Associates, LLC, manages and operates Jefferson Lab.
RAB also acknowledges support from the USDOE Early Career award, contract DE-SC0019229.

\appendix

\section{Justification replacing $\T_L$ with $\T$ \label{appendixA}}

Here we provide a justification of replacing $\T_L$ with $\T$ in Eq.~\eqref{eq:masterv2} up to errors that scale as $\mathcal{O}(e^{-mL})$. 

The key assumption made in Eq.~\eqref{eq:masterv2} is that the integral of the finite-volume Compton-amplitude $\mathcal{T}_L$ is equal to its infinite-volume counterpart, $\mathcal{T}_\infty$. It is certainly not true that $\mathcal{T}_L$ can be replaced with $\mathcal{T}_\infty$ in general. In fact, in general, these two quantities can differ by arbitrary amounts. Part of the point of Ref.~\cite{Briceno:2019opb} was to address this discrepancy for kinematics that may be accessed using lattice QCD. Here we use the findings of Ref.~\cite{Briceno:2019opb}, in addition to others. 

We are interested in volumes for which $\mathcal{O}(e^{-mL})$ errors can be neglected. In this region there are two other sources of large finite-volume effects. The first arise from new scales in the problem, such as the separation between the external currents. This, of course, is the focus of the main text of this work. The second arises from multi-particle states going on-shell. The manifestation of power-law finite-volume effects for two or more particle states is most famously associated with L\"uscher's work~\cite{Luscher:1991n1}.~\footnote{We point the reader to Refs.~\cite{Briceno:2017max, Hansen:2019nir} for recent reviews on the implementations and extensions of L\"uscher's formalism.}  As a result, all we need to argue is that the finite-volume effects associated with multi-particle states in $\mathcal{T}_L$ can be ignored in the integral.

\begin{figure}[t!]
\begin{center}
\includegraphics[width=\textwidth]{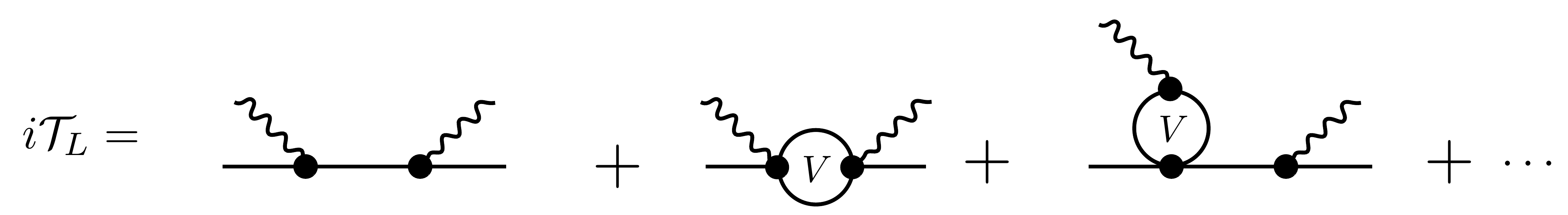}
\caption{Illustrative examples of the diagrams appearing in the definition of the finite-volume Compton-like amplitudes. The convention for the lines are the same as in Fig.~\ref{fig:iT}. The ``$V$" in the loops are meant to emphasize that the momenta of the intermediate particles must be discretized and therefore summed.}
\label{fig:iTL}
\end{center}
\end{figure}

Considering only contributions from two-particle states, we depict the two classes of potential finite-volume effects in Figure~\ref{fig:iTL} for $\mathcal{T}_L$. The first are the $s$- and $u$-channel two-particle loops. We illustrate the s-channel contributions explicitly in the second term on the righthand side of the equality, but leave the u-channel contributions implicit. The second class are those shown in the third term on the right-hand side, associated with time-like two-particle loops. The $s$- and $u$-channel finite volume corrections are related to $1\to2$ transitions~\cite{Briceno:2014uqa}, and these were the main focus of Ref.~\cite{Briceno:2019opb}. It was shown there that these corrections can be expressed to all orders in perturbation theory in terms of the infinite-volume $1\to2$ amplitude, the purely hadronic $2\to2$ scattering amplitude ($\mathcal{M}_2$), and a purely kinematic finite-volume function, $F$. This finite-volume function can be written in terms of the Riemann zeta functions. Similarly, one can write the finite-volume effects associated with the time-like diagrams to all orders in terms of $0\to 2$ transition amplitudes, $\mathcal{M}_2$, and the $F$ following formalism presented in Refs.~\cite{Briceno:2015csa, Meyer:2011um}.

%%%%%%%%%%%%%%%%%%%
%%%%%%%%%%%%%%%%%%%
%%%%%%%%%%%%%%%%%%%

Schematically, these finite-volume corrections are proportional to 
\begin{align}
\delta \mathcal{T}_L\propto F\frac{1}{1+\mathcal{M}_2\,F}.
\end{align} 
The poles of this function coincide with the finite-volume two-particle states, colloquially known as the L\"uscher poles. Assuming no UV cutoff, there is an infinite number of these poles, which we denote to be located at $M_n\geq2m$ where $n$ is a discrete integer enumerating the possible states. The residues of the poles depend on the finite-volume transition matrix elements, and we will compactly denote them as $c_n$. Therefore, the contributions to $\T_L$ coming from the L\"uscher poles can be written as,
\begin{align}
\delta \mathcal{T}_L(s)\propto \sum_n \frac{c_{n}}{s-M^2_n}.
\label{eq:Luscherprop}
\end{align} 
Here we are being schematic and assuming we are only interested in the $s$-dependent terms. Similar contributions can be written for $u$-channel diagrams and for those associated with the time-like form factors.

The finite-volume effects associated with the time-like form factor can immediately be identified to be of $\mathcal{O}(e^{-mL})$. This is because, as identified in the main body, the largest finite-volume effects come from the single-particle pole of the Compton-like amplitude. Given that the external states are on-shell, the residue of the finite-volume amplitude evaluated at the single-particle pole must depend on space-like virtualities. For these kinematics the finite-volume effects are $\mathcal{O}(e^{-mL})$.

This leaves us to consider the contributions from the L\"uscher poles in the $s$-/$u$-channel diagrams. To see the size of these, one can simply insert Eq.~\eqref{eq:Luscherprop} as a correction to the two-particle contributions of the $s$-/$u$-channel diagrams. This leads to contributions that are of the same kind as the ones considered in the main body of the text, with the mass of the light exchanged particle replaced with $M_n$. When the exchanged particle is of mass $m$ we know this leads to finite-volume errors of the order $\mathcal{O}(e^{-m|L-\xi|})$. Since the L\"uscher poles satisfy $M_n\geq 2m$, these corrections would scale at worst as $\mathcal{O}(e^{-2m|L-\xi|})$. For moderate values of $\xi < L/2$, these are subleading. 

%%%%%%%%%%%%%%%%%%%

\bibliography{bibi} 
\end{document}